\documentclass[pra,twocolumn,aps,amssymb,showpacs,superscriptaddress]{revtex4-1}

\usepackage{epsfig}
\usepackage{graphicx}
\usepackage{amsmath}
\usepackage{amssymb}
\usepackage{enumerate}
\usepackage{hyperref}
\newcommand{\sgn}{\text{sgn}}
\usepackage{color}
\definecolor{rosso}{rgb}{1,0,0}
\definecolor{verde}{rgb}{0,1,0}
\definecolor{blue}{rgb}{0,0,1}
\definecolor{amber}{rgb}{1.0, 0.75, 0.0}
\definecolor{amber(sae/ece)}{rgb}{1.0, 0.49, 0.0}
\definecolor{verdescuro}{rgb}{0,0.5,0.5}
\definecolor{rossoscuro}{rgb}{0.7,0.3,0}
\definecolor{bluscuro}{rgb}{0.3,0,0.7}
\definecolor{magenta}{rgb}{1,0,1}

\begin{document}

\title{Optimizing the proximity effect along the BCS-BEC crossover}

\author{V. Piselli}
\affiliation{School of Science and Technology, Physics Division, Universit\`{a} di Camerino, 62032 Camerino (MC), Italy}
\author{S. Simonucci}
\affiliation{School of Science and Technology, Physics Division, Universit\`{a} di Camerino, 62032 Camerino (MC), Italy}
\affiliation{INFN, Sezione di Perugia, 06123 Perugia (PG), Italy}
\author{G. Calvanese Strinati}
\email{giancarlo.strinati@unicam.it}
\affiliation{School of Science and Technology, Physics Division, Universit\`{a} di Camerino, 62032 Camerino (MC), Italy}
\affiliation{INFN, Sezione di Perugia, 06123 Perugia (PG), Italy}
\affiliation{CNR-INO, Istituto Nazionale di Ottica, Sede di Firenze, 50125 (FI), Italy}

\date{\today}

\begin{abstract}
The proximity effect, which arises at the interface between two fermionic superfluids with different critical temperatures, is examined with a non-local (integral) equation whose kernel contains information about the size of Cooper pairs that leak across the interface.
This integral approach avoids reference to the boundary conditions at the interface that would be required with a differential approach. 
The temperature dependence of the pair penetration depth on the normal side of the interface is determined over a wide temperature range, also varying the inter-particle coupling along the BCS-BEC crossover 
independently on both sides of the interface.
Conditions are found for which the proximity effect is optimized in terms of the extension of the pair penetration depth.
\end{abstract}


\maketitle

\section{Introduction} 
\label{sec:introduction}
\vspace{-0.3cm}

Recently, a novel approach for designing and tailoring entirely new classes of materials through ``proximity effects'' has ben suggested, which could overcome limitations inherent to more conventional (such as doping) methods \cite{PM-2018}.
Quite generally, proximity effects can rely on superconducting, magnetic, or topological properties.
Here, we consider theoretically the proximity effect that arises across the interface between two superconductors with different critical temperatures.
In this situation, the paired state in the superconductor at the left ($L$) of the interface, kept a temperature $T$ below its critical temperature $T_{c}^{L}$, leaks into the superconductor at the right ($R$) of the interface 
whose critical temperature $T_{c}^{R}$ is instead smaller than $T$. 
The novelty is that the (left) superconductor with higher temperature $T_{c}^{L}$ can be made to reach the so-called unitarity limit of the BCS-BEC crossover, where the size of the Cooper pairs becomes comparable with the inter-particle spacing \cite{Phys-Rep-2018}, in order to study the optimal conditions for the proximity effect to occur.

The above is a typical problem in inhomogeneous superconductivity, which can in principle be treated in terms of the Bogoliubov-de Gennes (BdG) equations with the inclusion of boundary effects 
\cite{deGennes-1964,deGennes-1966}.
As summarized in Ref.~\cite{Deutscher-1969}, however, most of the early knowledge about the proximity effect was gained in terms of the linearized Gor'kov equation for the gap parameter \cite{deGennes-1966}, which holds in the vicinity of the critical temperature. 
More recently, the BdG equations were used \cite{Klapwijk-2004} to demonstrate the connection between the proximity effect and the Andreev reflection \cite{Deutscher-2005}.

A characteristic quantity in the context of the proximity effect is the (temperature dependent) \emph{pair penetration depth} (or thickness of the leakage region on the normal side of the interface, which we shall refer to
as $\xi_{R}$ according to our reference geometry).
This quantity was estimated theoretically in Ref.~\cite{Kogan-1982} in terms of the Eilenberger formalism \cite{Eilenberger-1968} (or of its simplified Usadel version for the dirty limit \cite{Usadel-1970}), which is a ``quasi-classical'' approximation that greatly reduces the complexity of the Gor'kov equations by averaging out the fast oscillations (of the order of the Fermi wavelength) arising in the relative coordinate of Cooper pairs.
By this approach, in Ref.~\cite{Kogan-1982} it was possible to explore a wide interval of temperature which extends away from the immediate vicinity of the critical temperature, although still in the weak-coupling (BCS) regime of the superconducting coupling when a well-defined underlying Fermi surface is present.
 
Experiments could as well be directed at determining the temperature dependence of the pair penetration depth in the normal side, also in systems where the superconducting coupling may not be so weak,
along the lines of the original experimental work of Ref.~\cite{Polturak-1991}. 
In that work, critical-current measurements were performed in high-$T_{c}$ superconducting-normal-superconducting junctions, yielding an exponential dependence of the critical current on the thickness of the barrier 
which is a characteristic feature of the proximity effect.
In particular, from Fig.~4 of Ref.~\cite{Polturak-1991} one can identify a \emph{two-slope} dependence of the decay length in different temperature regimes (in the vicinity of $T_{c}^{R}$ and of $T_{c}^{L}$), a result which is in line with that obtained theoretically in Ref.~\cite{Kogan-1982} (albeit in the weak-coupling regime only).

This two-slope dependence of the pair penetration depth was qualitatively put in relation in Ref.~\cite{Palestini-2014} with the different temperature dependences that, in the normal phase above $T_{c}$, characterize
the healing length (due to \emph{inter}-pair correlations) and the pair coherence length (due to \emph{intra}-pair correlations). 
In Ref.~\cite{Palestini-2014}, however, the change of slope between these two lengths could be clearly identified only over a temperature range of the order of the Fermi temperature $T_{F}$, while in the proximity effect the pair penetration depth can be determined only over a much more limited temperature range since, in practice, $T_{c}^{L} \ll T_{F}$.
In addition, in Ref.~\cite{Palestini-2014} the two lengths were separately determined by two independent calculations, which could thus not identify them as the separate limiting values (close to and far away 
from the critical temperature $T_{c}$) of the same physical length.
In Ref.~\cite{Palestini-2014}, the study of these two lengths was systematically extended to the whole BCS-BEC crossover, throughout which the system evolves with continuity from large overlapping Cooper pairs (BCS regime) to dilute composite bosons (BEC regime)
 
With these premises, it appears highly desirable to study theoretically the proximity effect when either one of the fermionic systems at one side of the interface (or when even both of them) spans the BCS-BEC crossover up to the highest attainable temperature in the superfluid phase.
To this end, here we approach the problem in terms of a non-local (integral) equation for the gap parameter, which contains a kernel that depends on the gap parameter itself in a highly non-linear way and thus extends the linearized Gor'kov equation used originally in Ref.~\cite{deGennes-1964}, not only away from the vicinity of the critical temperature but also along the BCS-BEC crossover.
This integral equation was derived in Ref.~\cite{Simonucci-2014} by a double coarse-graining procedure applied to the BdG equations, which deals with the magnitude and phase of the gap parameter on a different footing (for this reason, the equation was referred to as the NLPDA equation with the acronym standing for Non Local Phase Density Approximation).
The properties and range of validity of the NLPDA equation were later discussed in Ref.~\cite{Simonucci-2017}, where an efficient practical method for finding its solution numerically was also provided. 
A key property, which renders the NLPDA equation ideally suited to deal with the proximity effect, is that the spatial extension of its kernel corresponds to the size of the Cooper pairs, for any coupling throughout the BCS-BEC crossover.

By making use of this approach, we will determine the pair penetration depth $\xi_{R}$ on the normal side of the interface over a wide temperature range and under quite different physical conditions on the two sides of the interface, thereby enabling us to identify the limiting behaviors (close to $T_{c}^{R}$ and to $T_{c}^{L}$) of this length in terms of a \emph{single} calculation.
In addition, by this approach we will have the flexibility of modelling the inter-particle coupling and the trapping potential in a physically smooth way across the interface that separates the left and right superconductors,
and we will avoid at the same time any reference to the boundary conditions at the interface \cite{deGennes-1964,deGennes-1966}.

The main results of our calculations are as follows: 

\noindent
(i) For given coupling at the left of the interface, the pair penetration depth $\xi_{R}$ at the right is found to increase (thereby amplifying the relevance of the proximity effect) when the bulk values $\Delta_{L}$ and $\Delta_{R}$, reached by the gap profile deep on the left and right of the interface, respectively, differ appreciably from each other.
This finding could also be used in reverse, in cases one would instead like to attenuate the occurrence of the proximity effect.

\noindent
(ii) When the coupling at the left of the interface is increased toward the unitary limit, such that the Cooper pair size decreases and becomes comparable with the inter-particle distance, 
the pair penetration depth $\xi_{R}$ at the right is found to decrease, too.
At the same time, however, there is an increase of the range of temperatures over which the proximity effect can occur.
Optimizing the proximity effect may thus require one to compromise between these two contrasting behaviors. 

\noindent
(iii) The temperature dependence of $\xi_{R}$ turns out to reproduce the behaviors in the vicinity of both $T_{c}^{R}$ and $T_{c}^{L}$ that were anticipated in Ref.~\cite{Kogan-1982} 
(although in that reference for the extreme weak-coupling limit only).
Our calculations extend these findings over a much wider coupling range along the BCS-BEC crossover.

\noindent
(iv) The pair penetration depth $\xi_{R}$ turns out to be essentially independent from the shape of the barrier, a feature which can be readily varied within the present approach. 

\noindent
(v) A ``negative'' proximity effect also occurs for the left superconductor with the higher temperature $T_{c}^{L}$, resulting in a marked depression of the gap profile which can extend far away from the interface.

In contrast to the present approach, more conventional treatments of the proximity effect in terms of the BdG equations \cite{Valls-2010} have largely focused on the region close to the interface (thereby not extracting the behavior of $\xi_{R}$), have described the interfacial scattering by a simple delta-function potential, have not pushed the calculation to the vicinity of the bulk transition temperature in the superconducting region, and, most importantly, have been limited only to the BCS (weak-coupling) limit of the BCS-BEC crossover.
However, consideration of the BCS-BEC crossover appears important not only for ultra-cold Fermi gases and nuclear systems \cite{Phys-Rep-2018}, but has recently acquired growing attention also in condensed matter
where experimental signatures of preformed Cooper pairing have been reported for Fe-based superconductors \cite{FeSe-2016}.
In addition, the conditions for the BCS-BEC crossover to occur could soon be purposely arranged in the emerging class of superconducting metamaterials \cite{metamaterials}, whereby the optimization of the proximity effect should prove especially relevant to the purpose.

The paper is organized as follows.
Section~\ref{sec:proximity-NLPDA} sets up the treatment of the proximity effect in terms of the NLPDA equation.
Section~\ref{sec:numerical-results} presents our numerical results for the profile of the gap parameter under a variety of circumstances, from which we are able to extract the temperature dependence of both the
pair penetration depth $\xi_{R}$ and coherence (healing) length $\xi_{L}$ at the right and left of the interface, respectively.
This information is then used for optimizing the proximity effect along the BCS-BEC crossover,
Section~\ref{sec:conclusions} gives our conclusions.
Finally, in Appendix~\ref{sec:appendix-A} a summary is given of the numerical procedure that solves the NLPDA equation in one dimension for the problem at hand.

\section{Proximity effect in terms of the (integral) NLPDA gap equation} 
\label{sec:proximity-NLPDA}

In this Section, we briefly recall the structure of the NLPDA equation, that was obtained in Ref.~\cite{Simonucci-2014} and further analyzed in Ref.~\cite{Simonucci-2017}, and 
reduce it to a one-dimensional form that corresponds to the proximity effect of interest when the gap parameter varies across the interface between two superconductors with different critical temperatures.
To this end, we shall need to specify the shape of the (smooth) variation of the coupling constant across the interface, as well as of the external potential which is required to keep the (left plus right) compound system at thermodynamic equilibrium.

\vspace{0.05cm}
\begin{center}
{\bf A. The NLPDA equation}
\end{center}

In Ref.~\cite{Simonucci-2014}, the following (integral) equation for the local gap parameter $\Delta(\mathbf{r})$
\begin{eqnarray}
- \frac{m}{4 \pi a_{F}} \, \Delta(\mathbf{r}) & = & \int \! d \mathbf{R} \,\, K(\mathbf{r}-\mathbf{R}|\mathbf{r}) \, \Delta(\mathbf{R}) 
\label{R-version-NLPDA-equation} \\
& = & \int \! \frac{d\mathbf{Q}}{\pi^{3}} \, e^{2 i \mathbf{Q} \cdot \mathbf{r}} \, K(\mathbf{Q}|\mathbf{r}) \, \Delta(\mathbf{Q})
\label{Q-version-NLPDA-equation}
\end{eqnarray}
\noindent
(referred to as the NLPDA equation) was obtained by a suitable coarse-graining procedure applied to the BdG equations.
(We set $\hbar = 1$ throughout.)
The kernel of this equation reads:
\begin{small}
\begin{equation}
K(\mathbf{Q}|\mathbf{r}) \! = \! \! \int \! \frac{d\mathbf{k}}{(2 \pi)^{3}}  \left\{ \frac{ 1 - 2 \, f_{F}(E_{+}(\mathbf{k};\mathbf{Q}|\mathbf{r}))}{2 E(\mathbf{k};\mathbf{Q}|\mathbf{r})} - \frac{m}{\mathbf{k}^{2}} \right\} 
\label{Kernel-Q} 
\end{equation}
\end{small}

\noindent
where $m$ is the fermion mass, $f_{F}(E) = \left( e^{E/(k_{B}T)} + 1 \right)^{-1}$ the Fermi function at temperature $T$ ($k_{B}$ being the Boltzmann constant),
\begin{footnotesize}
\begin{equation}
E_{\pm}(\mathbf{k};\mathbf{Q}|\mathbf{r}) =
\sqrt{\left( \frac{\mathbf{k}^{2}}{2m} + \frac{\mathbf{Q}^{2}}{2m} - \mu(\mathbf{r}) \right)^{2} + |\Delta(\mathbf{r})|^{2}}  \pm \frac{\mathbf{k}}{m} \cdot \mathbf{Q} \, ,
\label{definition-E-pm}
\end{equation}
\end{footnotesize}

\noindent
and $2 E(\mathbf{k};\mathbf{Q}|\mathbf{r}) = E_{+}(\mathbf{k};\mathbf{Q}|\mathbf{r}) + E_{-}(\mathbf{k};\mathbf{Q}|\mathbf{r})$.
In the above expression,  $\mu(\mathbf{r}) = \mu - V(\mathbf{r})$ is the local chemical potential in the presence of an external potential $V(\mathbf{r})$ 
and $|\Delta(\mathbf{r})|$ is the magnitude of the local gap parameter. 
(As they stand, the above expressions do not include the effects of a magnetic field.)
In addition, the kernel 
\begin{equation}
K(\mathbf{R}|\mathbf{r})  = \int \! \frac{d\mathbf{Q}}{\pi^{3}} \, e^{2 i \mathbf{Q} \cdot \mathbf{R}} \, K(\mathbf{Q}|\mathbf{r}) 
\label{Kernel-R}
\end{equation}
\noindent
in (real) $\mathbf{R}$-space results from Fourier transforming the kernel (\ref{Kernel-Q}) in (wave-vector) $\mathbf{Q}$-space.

The left-hand side of the NLPDA equation (in either form (\ref{R-version-NLPDA-equation}) or (\ref{Q-version-NLPDA-equation})) contains the scattering length $a_{F}$ for the two-fermion problem.
In terms of this quantity, one can form the dimensionless coupling parameter $(k_{F} a_{F})^{-1}$ that spans the BCS-BEC crossover \cite{Phys-Rep-2018}, 
where $k_{F} = (3 \pi^{2} n_{0})^{1/3}$ is the Fermi wave vector with (uniform) particle density $n_{0}$.
This parameter ranges from $(k_{F} a_{F})^{-1} \lesssim -1$ in the weak-coupling (BCS) regime when $a_{F} < 0$, to $(k_{F} a_{F})^{-1} \gtrsim +1$ in the strong-coupling (BEC) regime when 
$a_{F} > 0$, across the unitary limit when $|a_{F}|$ diverges (in practice, the ``crossover region'' $-1 \lesssim (k_{F} a_{F})^{-1} \lesssim +1$ is of most interest).

\vspace{0.05cm}
\begin{center}
{\bf B. The density equation}
\end{center}

The NLPDA integral equation (\ref{R-version-NLPDA-equation}) (or (\ref{Q-version-NLPDA-equation})) is highly non-linear in the gap parameter $\Delta$.
It thus generalises the linear integral equation adopted in Refs.~\cite{deGennes-1964,deGennes-1966,Deutscher-1969} to deal with the proximity effect, which (by construction) was valid  only in the vicinity of the superconducting transition $T_{c}^{L}$ when $\Delta$ is small (with respect to $k_{B} T_{c}^{L}$).
The NLPDA equation can then be applied for all temperatures in the superfluid phase and can also span the BCS-BEC crossover for arbitrary values of the coupling parameter 
$(k_{F} a_{F})^{-1}$, once it is supplemented by the density equation to determine the thermodynamic chemical potential $\mu$:
\begin{equation}
n(\mathbf{r}) \! = \! \! \int \! \frac{d\mathbf{k}}{(2 \pi)^{3}} \left\{1 - \frac{\xi(\mathbf{k}|\mathbf{r})}{E(\mathbf{k}|\mathbf{r})} \left[ 1 - 2 f(E(\mathbf{k}|\mathbf{r})) \right] \right\}
\label{density-equation}
\end{equation}
\noindent
where $\xi(\mathbf{k}|\mathbf{r}) = \frac{\mathbf{k}^{2}}{2m} - \mu(\mathbf{r})$ and $E(\mathbf{k}|\mathbf{r}) = \sqrt{\xi(\mathbf{k}|\mathbf{r})^{2} + |\Delta(\mathbf{r})|^{2}}$ \cite{Simonucci-2014}.
The expression (\ref{density-equation}) holds with a real gap parameter in the absence of currents.

\vspace{0.05cm}
\begin{center}
{\bf C. Variation of the coupling constant across the interface}
\end{center}

\begin{figure}[t]
\begin{center}
\includegraphics[width=7.8cm,angle=0]{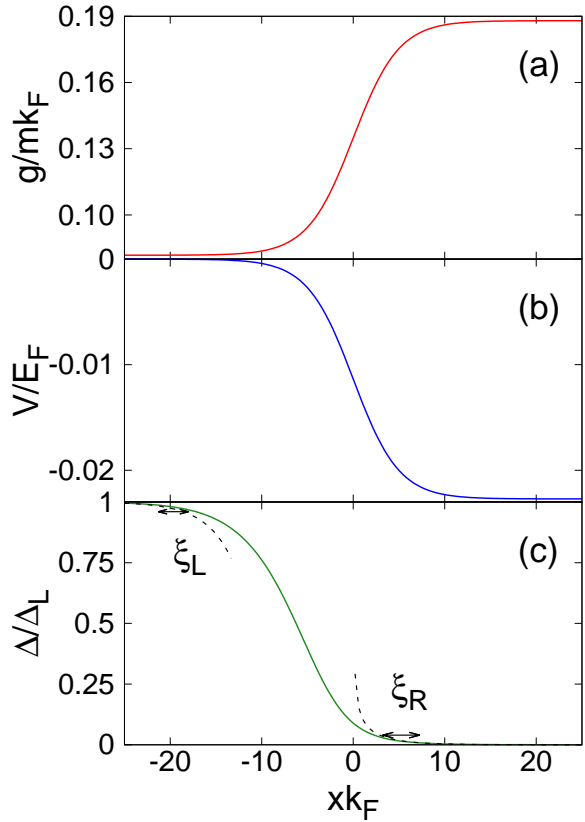}
\caption{(Color online) Characteristic spatial profiles of
             (a) the coupling constant $g(x)$ of Eq.~(\ref{model-function}) (in units of $m k_{F}$), 
             (b) the external potential $V(x)$ of Eq.~(\ref{V-model-dipole-layer}) (in units of the Fermi energy $E_{F}=k_{F}^{2}/(2m)$), 
             and (c) the gap parameter $\Delta(x)$ (in units of $\Delta_{L}$) obtained by solving the NLPDA equation in the form (\ref{NLPDA-eq-discretized}).
              For this example, we have taken $(k_{F} a_{F}^{L})^{-1} = - 1.0$ and $(k_{F} a_{F}^{R})^{-1} = - 2.34$ such that $T_{c}^{L}/T_{F} = 0.12$ and $T_{c}^{R}/T_{F}  = 0.015$, 
              $T/T_{F} = 0.09$ such that $T_{c}^{R} < T < T_{c}^{L}$, $\mu_{L}/E_{F} = 0.971$, $\mu_{R}/E_{F} = 0.993$, and $\Delta_{L}/E_{F} = 0.154$. 
              (Here $T_{F} = E_{F}/k_{B}$ is the Fermi temperature.) We have also taken $\sigma = 5 k_{F}^{-1}$ for the parameter that enters Eqs.~(\ref{model-function})-(\ref{V-model-dipole-layer}).
              The dashed lines in panel (c) represent the asymptotic expressions (\ref{asymptotic-behavior-Delta-R}) and (\ref{asymptotic-behavior-Delta-L}) from which $\xi_{R}$ and $\xi_{L}$ are extracted
              at given $T$.}
\label{Figure-1}
\end{center}
\end{figure}

In Refs.~\cite{deGennes-1964,deGennes-1966,Deutscher-1969} two different values of the inter-particle interaction were considered for the semi-infinite systems at the left ($L$) and right ($R$)
of the interface separating them at $x=0$.
By a similar token, here we attribute two different values of the coupling parameter $(k_{F} a_{F})^{-1}$ to the half-systems at the left and right of the interface, and assume
translational invariance in the $y$-$z$ plane parallel to the interface in such a way that both the external potential $V(x)$ and the gap parameter $\Delta(x)$ depend only on $x$.
To avoid too sharp a behaviour about $x=0$, it is convenient to smooth out the $x$-profile of the coupling parameter over a length $\sigma$ (of order of $k_{F}^{-1}$) by introducing the model function:
\begin{equation}
g(x) \equiv - \frac{m}{4 \pi a_{F}(x)} 
= \frac{1}{2} \left[ g_{R} + g_{L} + \left( g_{R} - g_{L} \right) G_{\sigma}(x) \right]
\label{model-function}
\end{equation}
\noindent
where $g_{L}=-m/(4 \pi a_{F}^{L})$ and $g_{R}=-m/(4 \pi a_{F}^{R})$ are the asymptotic values on the left and right sides of the interface, respectively.
For most calculations, we shall consider the function $G_{\sigma}(x)$ of the form
\begin{equation}
G_{\sigma}(x) = \tanh\left(\frac{x}{\sigma}\right) \, ;
\label{definition-profile-simple}
\end{equation}
\noindent
for the sake of comparison, however, we have sometimes utilized also the following function with compact support
\begin{eqnarray}
G_{\sigma}(x) & = & \tanh \! \left( \frac{\frac{x}{\sigma}}{\sqrt{1 - \left(\frac{x}{\sigma}\right)^{2}}} \right) \hspace{0.3cm} \left[ \frac{|x|}{\sigma} \le 1 \right]
\nonumber \\
& = & \sgn{\left(\frac{x}{\sigma}\right)} \hspace{2.2cm} \left[ \frac{|x|}{\sigma} \ge 1 \right]\, .
\label{definition-profile-complex}
\end{eqnarray}
A typical profile of $g(x)$ with the choice (\ref{definition-profile-simple}) is shown in Fig.~\ref{Figure-1}(a).

\begin{center}
{\bf D. Choice of the external potential}
\end{center}

The potential $V(x)$, that enters the NLPDA equation (\ref{R-version-NLPDA-equation}) (or (\ref{Q-version-NLPDA-equation})) and the density equation (\ref{density-equation}) through the local chemical potential 
$\mu(x)$, can be modelled in several ways, depending on the experimental conditions one is after.
Here, we consider the following choice for $V(x)$.

We assume that the system at the left (right) of the interface extends to $-\infty$ ($+\infty$), such that away from the interface in the bulk region it approaches a homogeneous 
superconductor with coupling $(k_{F}a_{F}^{L})^{-1}$ ($(k_{F}a_{F}^{R})^{-1}$) and bulk (asymptotic) value $\mu_{L}$ ($\mu_{R}$) of the chemical potential.
For simplicity, we further assume that the density has the same bulk (asymptotic) value $n_{0}= k_{F}^{3}/(3 \pi^{2})$ on both sides of the interface.
At a given temperature, this corresponds to the value $n(-\infty)$ obtained from Eq.(\ref{density-equation}) with chemical potential $\mu_{L}$ and $\Delta(-\infty)=\Delta_{L}$, 
and to the value $n(+\infty)$ obtained from Eq.(\ref{density-equation}) with chemical potential $\mu_{R}$ and $\Delta(+\infty)=\Delta_{R}$.
However, since at equilibrium the thermodynamic chemical potential $\mu$ must maintain the same value across the whole system, the situation can be kept thermodynamically stable only in the presence of an external potential $V(x)$, which makes the ``local'' chemical potential $\mu(x) = \mu - V(x)$ entering Eq.(\ref{density-equation}) to interpolate smoothly between the asymptotic values $\mu_{L}$ and $\mu_{R}$.
In analogy with Eq.~(\ref{model-function}), we write:
\begin{equation}
V(x) = \mu - \frac{1}{2} \left[ \left( \mu_{R} + \mu_{L} \right) + \left( \mu_{R} - \mu_{L} \right) G_{\sigma}(x) \right]
\label{V-model}
\end{equation}
\noindent
whereby $V(-\infty) = \mu - \mu_{L}$ and $V(+\infty) = \mu - \mu_{R}$.
At a given temperature, the arbitrariness on the value of $\mu$ can be eliminated by fixing $\mu = \mu_{L}$, which corresponds to a homogeneous system with density $n_{0}$ and coupling $(k_{F}a_{F}^{L})^{-1}$.
In this way, $V(-\infty) = 0$ and the expression (\ref{V-model}) reduces to the form:
\begin{equation}
V(x) = \frac{1}{2} \left( \mu_{L} - \mu_{R} \right) \, \left[ 1 \, + \, G_{\sigma}(x) \right] 
\label{V-model-dipole-layer}
\end{equation}
\noindent
which depends on temperature through the temperature dependence of both $\mu_{L}$ and $\mu_{R}$.
In particular, when $T_{c}^{R} < T_{c}^{L}$ one expects $\mu_{L} < \mu_{R}$, such that $V(x) \le 0$ from Eq.(\ref{V-model-dipole-layer}).
A typical profile of $V(x)$ is shown in Fig.~\ref{Figure-1}(b).
The potential $V(x)$ thus acts as a ``barrier'' that effectively prevents the particles from flowing from the right toward the left of the interface, while trying to take advantage of the smaller local value of the chemical potential.
Accordingly, close to the interface one expects the local density $n(x)$ to somewhat deviate from its bulk value $n_{0}$, possibly leading to a depression on one side and to an enhancement on the other side of the interface.
In a condensed-matter system, this situation would correspond to the presence of an electrostatic dipole layer across the boundary surface \cite{Jackson-2009}. 

\vspace{0.05cm}
\begin{center}
{\bf E. Solution of the NLPDA equation across the interface}
\end{center}

Under the above circumstances, the gap parameter in $\mathbf{Q}$-space that enters the right-hand side of Eq.(\ref{Q-version-NLPDA-equation}) reduces to the form:
\begin{equation}
\Delta(\mathbf{Q}) = \pi^{2} \delta(Q_{y}) \delta(Q_{z}) \Delta(Q_{x}) \, .
\label{one-dimensional-gap-parameter}
\end{equation}
\noindent
Correspondingly, the NLPDA equation (\ref{Q-version-NLPDA-equation}) simplifies as follows:
\begin{equation}
g(x) \Delta(x) = \int_{- \infty}^{+\infty} \! \frac{dQ}{\pi} \, e^{2iQx} \, K(|Q||x) \, \Delta(Q)
\label{1D-NLPDA-equation}
\end{equation}
\noindent
with the notation of Eq.(\ref{model-function}) and where we have set $Q_{x} \rightarrow Q$ in Eq.(\ref{1D-NLPDA-equation}) to shorten the notation.
Note that in the expression (\ref{1D-NLPDA-equation}) we have emphasized the fact that the kernel $K$ depends on the magnitude $|Q|$ of $Q$.

The integral equation (\ref{1D-NLPDA-equation}) can be solved using general method developed in Appendix B of Ref.~\cite{Simonucci-2017}, where the Fourier transform of a function with a given spatial symmetry 
in D-dimensions was calculated in terms of the eigen-functions of the harmonic oscillator.
In Appendix~\ref{sec:appendix-A} below this method is further adapted to the present 1D case, whereby the gap parameter $\Delta(x)$ is neither symmetric nor antisymmetric across the interface at $x=0$, and, 
in addition, the coupling parameter $g(x)$ depends on $x$.
The end result is the following discretized expression of the 1D-NLPDA integral equation (\ref{1D-NLPDA-equation}) \cite{footnote-I}
\begin{widetext}
\begin{equation}
g \left(\! \frac{x_{j}}{\sqrt{2} \lambda} \! \right) \Delta \left(\! \frac{x_{j}}{\sqrt{2} \lambda} \! \right) = \frac{1}{y_{j}} \sum_{n=0}^{N-1} \sum_{j'=1}^{N} \, i^{n} \, S^{T}_{jn} \, S_{nj'} \, 
K \! \left(\! \frac{\lambda |x_{j'}|}{\sqrt{2}} \Big| \frac{x_{j}}{\sqrt{2} \lambda} \! \right)
\, \sum_{n'=0}^{N-1} \sum_{j''=1}^{N} \, (-i)^{n'} \, S^{T}_{j'n'} \, S_{n'j''} \, y_{j''} \, \Delta \left(\! \frac{x_{j''}}{\sqrt{2} \lambda} \! \right) \, .
\label{NLPDA-eq-discretized}
\end{equation}
\end{widetext}
\noindent
In this expression: 
(i) $\frac{x_{j}}{\sqrt{2} \lambda}$ refers to values of $x$ in real space and $\frac{\lambda x_{j}}{\sqrt{2}}$ to values of $Q$ in wave-vector space;
(ii) the points $\{ x_{j};j=1,\cdots,N \}$ correspond to the zeros of the (normalized) Hermite polynomial $\mathcal{H}_{N}(x)$ (cf. Eq.~{\ref{big-eigenvalue-problem}));
(iii) the matrix elements of the orthogonal matrix $S$ are given by Eq.~(\ref{definition-matrix_S-y});
(iv) the positive definite weights $w_{j}$ are obtained by the normalization condition (\ref{normalization-condition}) for the eigenvectors of the eigenvalue problem (\ref{big-eigenvalue-problem});
and (v) the quantity $y_{j}$ is given by Eq.~(\ref{definition-matrix_S-y}).
The (positive) parameter $\lambda$ is meant to add extra flexibility to the numerical calculations.
The number of points $N$ in the two ($x$ and $Q$) meshes and the parameter $\lambda$ can be varied to achieve optimal convergence of the Fourier transforms, from $\Delta(x)$ to $\Delta(Q)$ and viceversa.
A typical profile for $\Delta(x)$ obtained in this way is shown in Fig.~\ref{Figure-1}(c).

The discretized version (\ref{NLPDA-eq-discretized}) of the 1D-NLPDA equation is solved until self-consistency is achieved, by following closely the prescriptions discussed in Appendix B of Ref.~\cite{Simonucci-2017}. 
In practice, the values of the gap $\Delta(x)$ are explicitly calculated over a coarse mesh of $M$ points (we have taken $M = 350$ in most calculations). 
A numerical interpolation is then used to generate the $N (> M)$ values of $\Delta(x)$ needed in Eq.~(\ref{NLPDA-eq-discretized}) (typically, $N = 30 M$ proves sufficient). 
This interpolation is required to avoid unwanted oscillations of small wavelengths which would be generated otherwise. 
Finally, the value of the parameter $\lambda$ entering Eq.~(\ref{NLPDA-eq-discretized}) is chosen in such away that the $N$ zeros of the Hermite polynomials extend over a spatial range 
which is three times wider than that covered by the $M$ points utilized for the profile of $\Delta(x)$ (typically, we have taken $\lambda k_{F}  = 10$).
For the needs of the present paper, the self-consistent solution of Eq.~ (\ref{NLPDA-eq-discretized}) has been achieved in a few hundreds cases.

\section{Numerical results} 
\label{sec:numerical-results}

\begin{figure}[h]
\begin{center}
\includegraphics[width=7.8cm,angle=0]{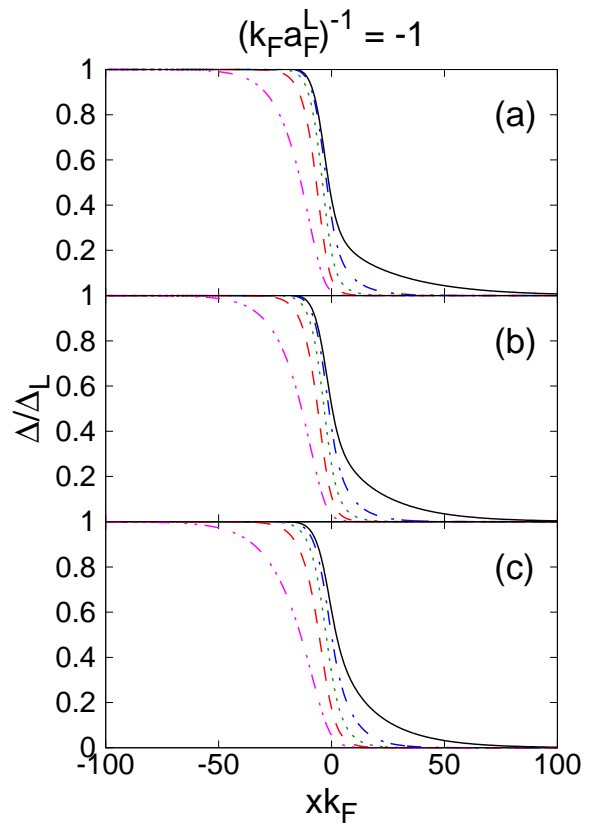}
\caption{(Color online) Gap profiles $\Delta(x)/\Delta_{L}$ for $(k_{F} a_{F}^{L})^{-1} = -1.0$ when (a) $T_{c}^{L}/T_{c}^{R} = 8$, (b) $T_{c}^{L}/T_{c}^{R} = 4$, and (c) $T_{c}^{L}/T_{c}^{R} = 2$.
              The various curves refer to different temperatures chosen according to the expression (\ref{temperature-interpolated}), where
              $\nu = 0.05$ (full line),
              $\nu = 0.25$ (dashed-dotted line),
              $\nu = 0.50$ (dotted line),
              $\nu = 0.75$ (dashed line),
              $\nu = 0.95$ (dashed-double-dotted line).}
\label{Figure-2}
\end{center}
\end{figure} 
The numerical solution of the integral equation (\ref{NLPDA-eq-discretized}) has been performed in several cases, by varying the coupling constants $(k_{F} a_{F}^{L})^{-1}$ and $(k_{F} a_{F}^{R})^{-1}$ at the left and right of the interface as well as the width $\sigma$ of the separating barrier.
In particular, we have considered the values $(k_{F} a_{F}^{L})^{-1} = (-1.0,0.0)$ such that $T_{c}^{L}/T_{F} = (0.12,0.50)$, and we have correspondingly adapted the value of $(k_{F} a_{F}^{R})^{-1}$ such that
$T_{c}^{L}/T_{c}^{R} = (8,4,2)$ \cite{footnote-II}.
In addition, we have taken $k_{F} \sigma = (2.5,5.0,10.0)$ for the choice (\ref{definition-profile-simple}) and $k_{F} \sigma = 5.0$ for the choice (\ref{definition-profile-complex}) of $G_{\sigma}(x)$.
This wide choice of input parameters will enable us to draw some definite conclusions about the way the proximity effect can be optimized (or, in reverse, depressed).

\begin{center}
{\bf A. Profile of the gap parameter across the interface under various circumstances}  
\end{center}

The basic results of the present calculation are represented by the gap profiles $\Delta(x)$ across the interface.
Several examples of these profiles are shown in Figs.~\ref{Figure-2} and \ref{Figure-3} for $(k_{F} a_{F}^{L})^{-1} = -1.0$ and $(k_{F} a_{F}^{L})^{-1} = 0.0$, respectively,
with the choice $k_{F}\sigma = 5.0$ for the barrier (\ref{definition-profile-simple}).
In each figure, the three panels refer to the cases (a) $T_{c}^{L}/T_{c}^{R} = 8$, (b) $T_{c}^{L}/T_{c}^{R} = 4$, and (c) $T_{c}^{L}/T_{c}^{R} = 2$.
In each panel, several temperatures are further considered according to the expression
\begin{equation}
\frac{T}{T_{c}^{L}} = \nu + (1 - \nu) \, \frac{T_{c}^{R}}{T_{c}^{L}} \hspace{1.0cm} ( 0 \le \nu \le 1) \, ,
\label{temperature-interpolated}
\end{equation}
\noindent
such that $T = T_{c}^{R}$ for $\nu = 0$ and $T = T_{c}^{L}$ for $\nu = 1$.
In particular, in Figs.~\ref{Figure-2} and \ref{Figure-3} we have chosen the values $\nu = (0.05,0.25,0.50,0.75,0.95)$.
\begin{figure}[t]
\begin{center}
\includegraphics[width=7.8cm,angle=0]{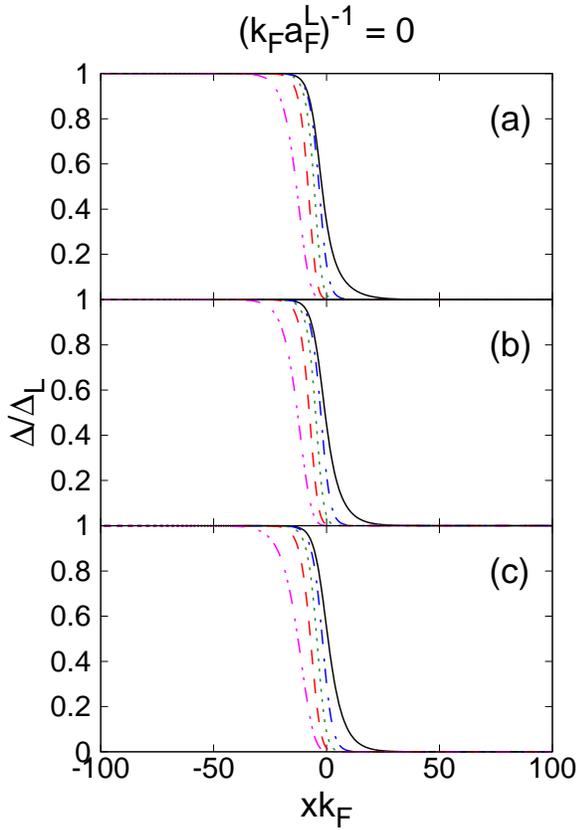}
\caption{(Color online) Gap profiles $\Delta(x)/\Delta_{L}$ for $(k_{F} a_{F}^{L})^{-1} = 0.0$ when (a) $T_{c}^{L}/T_{c}^{R} = 8$, (b) $T_{c}^{L}/T_{c}^{R} = 4$, and (c) $T_{c}^{L}/T_{c}^{R} = 2$.
              Conventions for the various curves are the same of Fig.~\ref{Figure-2}.}
\label{Figure-3}
\end{center}
\end{figure} 

It is also interesting to compare the gap profiles $\Delta(x)$ for different shapes of the barrier.
This is done in Fig.~\ref{Figure-4}, where several values of the barrier width $\sigma$ are used for the choice (\ref{definition-profile-simple}), and a single value of $\sigma$ is
considered for the choice (\ref{definition-profile-complex}).
In particular, panel (b) of Fig.~\ref{Figure-4} shows that, on the right side of the interface, the gap profile is essentially independent from the shape of the barrier.
This result gives us confidence that the values of the pair penetration depth $\xi_{R}$, that we will extract from $\Delta(x)$ for $x>0$ to characterize the proximity effect, will not appreciably depend 
on a specific choice of the barrier.
For the sake of definiteness, in what follows we will thus limit ourselves to consider a barrier specified by the form (\ref{definition-profile-simple}) with $k_{F} \sigma = 5.0$.
\begin{figure}[t]
\begin{center}
\includegraphics[width=7.5cm,angle=0]{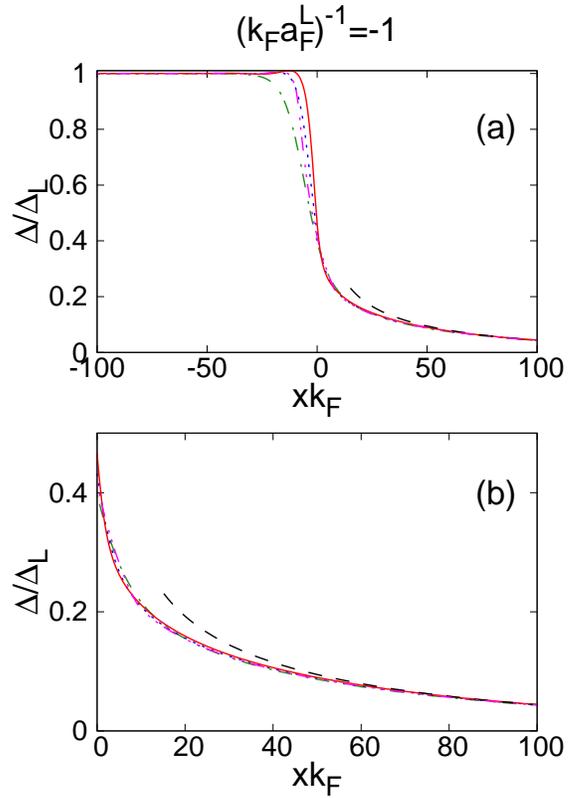}
\caption{(Color online) The gap profile $\Delta(x)/\Delta_{L}$ for $(k_{F} a_{F}^{L})^{-1} = -1.0$, $T_{c}^{L}/T_{c}^{R} = 8$, and $T = 1.04 T_{c}^{R}$ is shown for several values of the barrier width 
              $\sigma$ of Eq.~(\ref{definition-profile-simple}):
              $k_{F}\sigma = 2.5$ (full line), $k_{F}\sigma = 5.0$ (dotted line), $k_{F}\sigma = 10.0$ (dashed-dotted line).
              Also reported are the results for the choice (\ref{definition-profile-complex}) with $k_{F}\sigma = 5.0$ (dashed-double-dotted line) and of the fitting (\ref{xi-R-L-fitting-1}) 
              (dashed line) - see below.
              Panel (a) shows the whole profile of $\Delta(x)$ both at the left and right of the interface, while panel (b) focuses on the right side of the interface from which the pair penetration depth $\xi_{R}$ of
              interest is extracted.}
\label{Figure-4}
\end{center}
\end{figure} 

\begin{center}
{\bf B. Asymptotic behavior of the gap parameter \\ on both sides of the interface}
\end{center}

Out of the numerical results for $\Delta(x)$ like those reported in Figs.~\ref{Figure-2} and \ref{Figure-3}, one can extract both the pair penetration depth $\xi_{R}$ and the coherence (healing) length $\xi_{L}$ according to the following procedure.
At a given temperature $T$, we fit the behavior of $\Delta(x;T)$ for $k_{F} x \gg 1$ through the expression
\begin{equation}
\Delta(x;T) \sim \frac{\gamma_{R}(T) \, e^{-x/\xi_{R}(T)} }{x^{\mathcal{D}-2+\eta}} + \Delta_{R}(T) \, ,
\label{asymptotic-behavior-Delta-R} 
\end{equation}
\noindent
while for $k_{F} x \ll -1$ we make use of the specular expression
\begin{equation}
\Delta(x;T) \sim \Delta_{L}(T) - \frac{\gamma_{L}(T) \, e^{x/\xi_{L}(T)}}{|x|^{\mathcal{D}-2+\eta}}  \, .
\label{asymptotic-behavior-Delta-L} 
\end{equation}
\noindent 
In all fits we that have performed, it turns out that the optimal value of $\mathcal{D} + \eta $ is $2.5$.
We have then set $\eta = 0$ and interpreted $\mathcal{D} = 2.5$ as an ``effective'' dimensionality, which is intermediate between $D=2$ of the planar boundary surface separating the left ($L$) and right ($R$) superconductors and $D=3$ of the space 
in which this surface is embedded.
Note that the expressions (\ref{asymptotic-behavior-Delta-R}) and (\ref{asymptotic-behavior-Delta-L}) correspond to the generic behavior of the correlation function for the order parameter in a homogeneous 
medium \cite{LeBellac-1995}, and are here recovered by the spatial behaviour of the order parameter itself for the inhomogeneous problem we are considering \cite{footnote-III}.
In the expressions (\ref{asymptotic-behavior-Delta-R}) and (\ref{asymptotic-behavior-Delta-L}), note also the presence of the temperature dependent (and positive definite) pre-factors $\gamma_{R}(T)$ and 
$\gamma_{L}(T)$, which are needed for obtaining accurate fits of the asymptotic gap profiles.

\begin{center}
{\bf C. Optimizing the proximity effect in terms of $\xi_{R}$}
\end{center}

\begin{figure}[t]
\begin{center}
\includegraphics[width=8.5cm,angle=0]{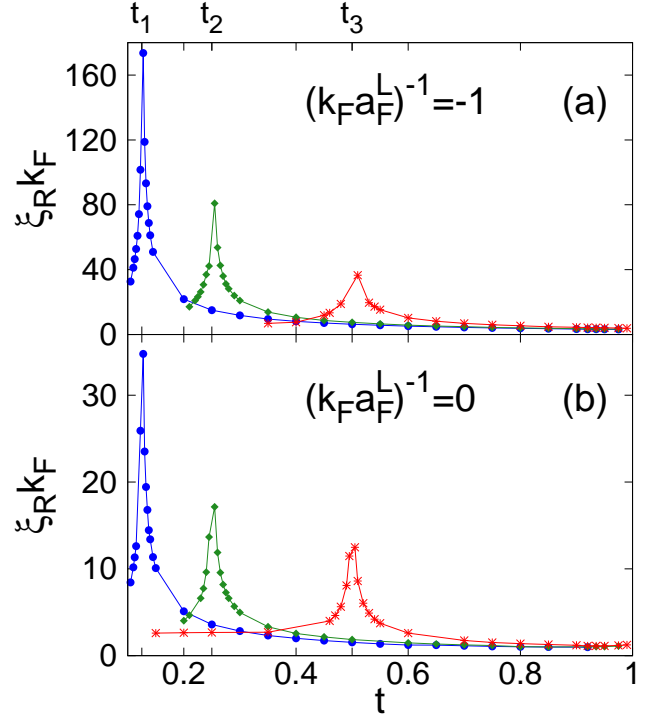}
\caption{(Color online) The pair penetration depth $\xi_{R}$ (in units of $k_{F}^{-1}$) is shown vs the reduced temperature $t=T/T_{c}^{L}$ for the coupling values 
              (a) $(k_{F} a_{F}^{L})^{-1} = -1.0$ and (b) $(k_{F} a_{F}^{L})^{-1} = 0.0$. In each panel, three cases are reported for $T_{c}^{R}/T_{c}^{L} = t_{1} = 1/8$ (dots), $T_{c}^{R}/T_{c}^{L} = t_{2} = 1/4$ (diamonds),
              and $T_{c}^{R}/T_{c}^{L} = t_{3} = 1/2$ (stars). (The lines are guides to the eye.)}
\label{Figure-5}
\end{center}
\end{figure} 

Figure~\ref{Figure-5} shows the results for $\xi_{R}(T)$ obtained from a fit of the form (\ref{asymptotic-behavior-Delta-R}),
for the coupling values $(k_{F} a_{F}^{L})^{-1} = -1.0$ (upper panel) and $(k_{F} a_{F}^{L})^{-1} = 0.0$ (lower panel).
In each panel, three different cases are reported with $T_{c}^{R}/T_{c}^{L} = 1/8$, $T_{c}^{R}/T_{c}^{L} = 1/4$, and $T_{c}^{R}/T_{c}^{L} = 1/2$.
For both couplings of the left superconductor, it is seen that $\xi_{R}$ attains larger values as soon as the coupling of the right superconductor differs appreciably from that of the left superconductor.
This implies that the relevance of the proximity effect is amplified when the bulk values $\Delta_{L}$ and $\Delta_{R}$ differ appreciably from each other, a criterion that could be exploited in practice
to optimize the spatial extension of the proximity effect.
For the sake of example, typical values of $\Delta_{L}$ and $\Delta_{R}$ at zero temperature are reported in Table I for the cases of interest.

\begin{table}
\begin{center}
\begin{tabular}{ | c | c | c | c | c | c |}
    \hline
    $(k_{F}a_{F}^{L})^{-1}$ & $(k_{F}a_{F}^{R})^{-1}$ & $T_{c}^{L}/T_{F}$ & $T_{c}^{R}/T_{F}$ & $\Delta_{L}/E_{F}$ & $\Delta_{R}/E_{F}$  \\ \hline \hline
    -1.0 & -2.36 & 0.12 & 0.015 & 0.20  & 0.026  \\ \hline
    -1.0 & -1.92 & 0.12 & 0.030 & 0.20  & 0.053  \\ \hline
    -1.0 & -1.48 & 0.12 & 0.060 & 0.20  & 0.104  \\ \hline
     0.0 & -1.45 & 0.50 & 0.063 & 0.69  & 0.108  \\ \hline
     0.0 & -1.00 & 0.50 & 0.125 & 0.69  & 0.208  \\ \hline
     0.0 & -0.53 & 0.50 & 0.250 & 0.69  & 0.388  \\ \hline
\end{tabular}
\caption{Values of $\Delta_{L}$ and $\Delta_{R}$ (in units of the Fermi energy $E_{F}$) at zero temperature for the couplings of interest \cite{footnote-II}.}
\end{center}
\end{table}

When comparing the sets of values for $\xi_{R}$ that correspond to $(k_{F} a_{F}^{L})^{-1} = -1.0$ and $(k_{F} a_{F}^{L})^{-1} = 0.0$, as reported in panels (a) and (b) of Fig.~\ref{Figure-5}, 
respectively, one notices that those for $(k_{F} a_{F}^{L})^{-1} = -1.0$ result always larger than those for $(k_{F} a_{F}^{L})^{-1} = 0.0$.
This is in line with the fact that, for a homogeneous system, the Cooper pair size $\xi_{\mathrm{pair}}$ is smaller for $(k_{F} a_{F}^{L})^{-1} = 0.0$ (where $k_{F} \xi_{\mathrm{pair}} = 1.1$) than for 
$(k_{F} a_{F}^{L})^{-1} = -1.0$ (where $k_{F} \xi_{\mathrm{pair}} = 3.4$) \cite{Pistolesi-1994}, such that the leakage region on the normal side of the interface associated with the proximity effect should 
correspondingly be smaller.
On the other hand, one should also recall that, in absolute value, the range of temperatures over which the proximity effect can occur increases from $(k_{F} a_{F}^{L})^{-1} = 0.0$ to 
$(k_{F} a_{F}^{L})^{-1} = -1.0$, to the extent that the corresponding critical temperature $T_{c}^{L}$ is larger when $(k_{F} a_{F}^{L})^{-1} = 0.0$.
Optimizing the proximity effect may thus require one to compromise between these two contrasting behaviors, depending on the physical circumstances of interest.

\begin{center}
{\bf D. Limiting behaviors for the temperature dependence of $\xi_{R}$}
\end{center}

The numerical results for $\xi_{R}(T)$ reported in Fig.~\ref{Figure-5} can be further analyzed for temperatures close enough to $T_{c}^{R}$ and $T_{c}^{L}$. 
To this end, we resort to the analytic results that were obtained in Ref.~\cite{Kogan-1982} for the extreme weak-coupling (BCS) limit only and utilize them also for stronger couplings reaching the unitary limit,
in order to fit the temperature dependence of $\xi_{R}$ and $\xi_{L}$ obtained above out of the expressions (\ref{asymptotic-behavior-Delta-R}) and (\ref{asymptotic-behavior-Delta-L}).
Accordingly, we represent the temperature dependence of $\xi_{R}$ both close to $T_{c}^{R}$ and $T_{c}^{L}$ (as well as of $\xi_{L}$ close to $T_{c}^{L}$) in the following way:
\begin{eqnarray}
\xi_{R}(T) & = & \frac{A_{R}^{(+)}}{\sqrt{T - T_{c}^{R}}} \hspace{0.5cm} [T \gtrsim T_{c}^{R}]
\label{xi-R-L-fitting-1}  \\
\xi_{R}(T) & = & \frac{A_{R}^{(-)}}{\sqrt{T_{c}^{R} - T}} \hspace{0.5cm} [T \lesssim T_{c}^{R}]
\label{xi-R-L-fitting-2} \\
\xi_{R}(T) & = & \frac{B_{R}}{T} \hspace{1.3cm} [T_{c}^{R} \ll T \lesssim T_{c}^{L}]
\label{xi-R-L-fitting-3} \\
\xi_{L}(T) & = & \frac{A_{L}^{(-)}}{\sqrt{T_{c}^{L} - T}} \hspace{0.5cm} [T \lesssim T_{c}^{L}]
\label{xi-R-L-fitting-4}
\end{eqnarray}
\noindent
where ``$\gtrsim$'' and ``$\lesssim$'' signify ``in the vicinity of'' and ``$\ll$'' signifies ``well above than''.

\begin{figure}[t]
\begin{center}
\includegraphics[width=7.8cm,angle=0]{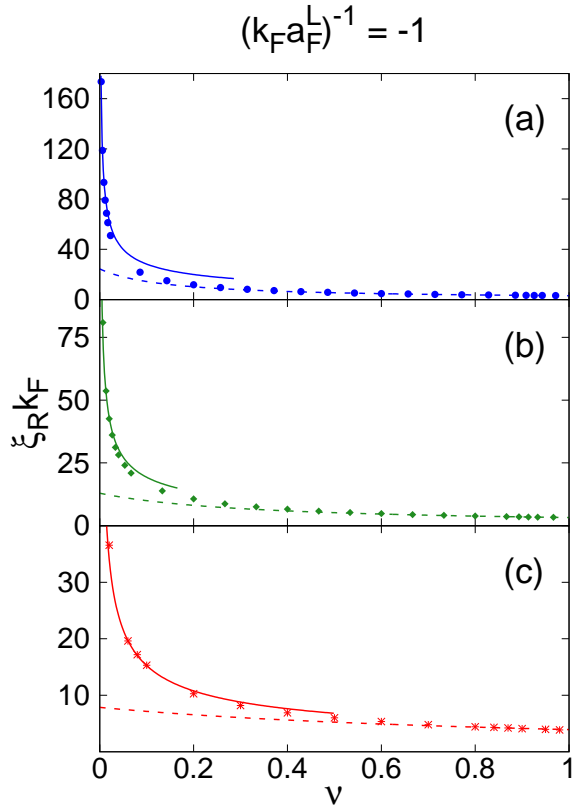}
\caption{(Color online) The pair penetration depth $\xi_{R}$ (in units of $k_{F}^{-1}$) is reported vs the variable $\nu$ of Eq.~(\ref{temperature-interpolated}) for $(k_{F} a_{F}^{L})^{-1} = -1.0$,
                                     in the three cases when $T_{c}^{L}/T_{c}^{R} = 8$ (dots - upper panel), $T_{c}^{L}/T_{c}^{R} = 4$ (diamonds - middle panel), and  $T_{c}^{L}/T_{c}^{R} = 2$ (stars - lower panel).
                                     In addition, fits to these symbols are obtained with the expressions (\ref{xi-R-L-fitting-1}) close to $T_{c}^{R}$ (full lines) and (\ref{xi-R-L-fitting-3}) close to $T_{c}^{L}$ (dashed lines).}
\label{Figure-6}
\end{center}
\end{figure} 
\begin{figure}[h]
\begin{center}
\includegraphics[width=7.8cm,angle=0]{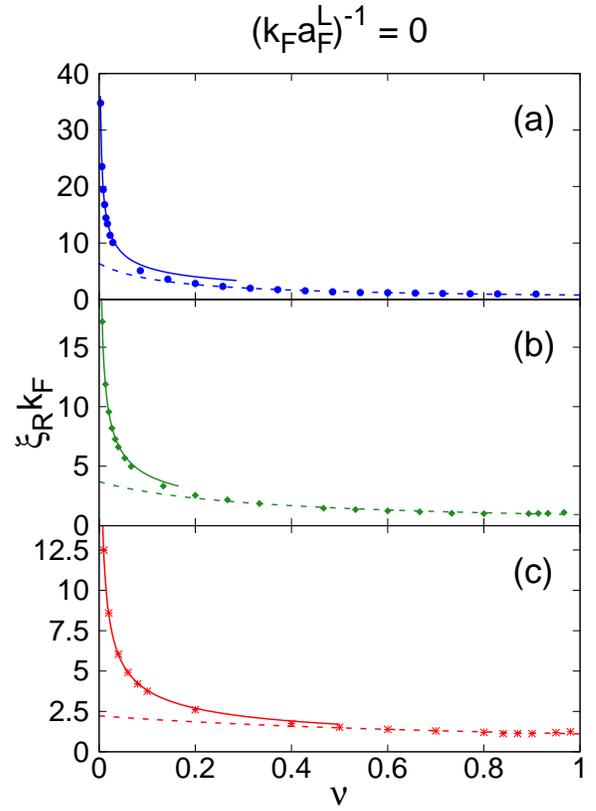}
\caption{(Color online) Pair penetration depth $\xi_{R}$ (in units of $k_{F}^{-1}$) vs the variable $\nu$ of Eq.~(\ref{temperature-interpolated}) for $(k_{F} a_{F}^{L})^{-1} = 0.0$.
                                    Conventions and symbols are the same of Fig.~\ref{Figure-6}.}
\label{Figure-7}
\end{center}
\end{figure} 

The results of these fits for $\xi_{R}$ are shown in Fig.~\ref{Figure-6} for $(k_{F} a_{F}^{L})^{-1} = -1.0$ and in Fig.~\ref{Figure-7} for $(k_{F} a_{F}^{L})^{-1} = 0.0$,
where in each case $T_{c}^{L}/T_{c}^{R} = 8$ (upper panel), $T_{c}^{L}/T_{c}^{R} = 4$ (middle panel), and $T_{c}^{L}/T_{c}^{R} = 2$ (lower panel).
Note that, to draw these three different cases over the same horizontal scale, we have identified the reduced temperature $T/T_{c}^{L}$ in terms of the variable $\nu$ of Eq.~(\ref{temperature-interpolated}),
such that $T = T_{c}^{R}$ when $\nu = 0$ and $T = T_{c}^{L}$ when $\nu = 1$. 
In each case, the numerical values for $\xi_{R}$ (symbols) are fitted close to $T_{c}^{R}$ via the expression (\ref{xi-R-L-fitting-1}) (full lines) and close to $T_{c}^{L}$ via the expression (\ref{xi-R-L-fitting-3})
(dashed lines).
The values of all coefficients entering the expressions (\ref{xi-R-L-fitting-1})-(\ref{xi-R-L-fitting-4}) are reported in Table II for all cases considered in Figs.~\ref{Figure-6} and \ref{Figure-7}.
\begin{table}
\begin{center}
\begin{tabular}{ | c | c | c | c | c | c | c | c |}
    \hline
    $(k_{F}a_{F}^{L})^{-1}$ & $(k_{F}a_{F}^{R})^{-1}$ & $T_{c}^{L}/T_{F}$ & $T_{c}^{R}/T_{F}$ & $A_{R}^{(+)}$ & $A_{R}^{(-)}$ & $B_{R}$ & $A_{L}^{(-)}$  \\ \hline \hline
    -1.0 & -2.36  & 0.12  & 0.015  & 192.7  & 119.7   & 198.9   & 5.7   \\ \hline
    -1.0 & -1.92  & 0.12  & 0.030  &   60.9   & 38.5    &   74.7   & 5.6   \\ \hline
    -1.0 & -1.48  & 0.12  & 0.060  &   19.7   & 15.4    &   32.1   & 5.7   \\ \hline
     0.0 & -1.45  & 0.50  & 0.063  &   19.1   & 14.5    &   26.1   & 1.2    \\ \hline
     0.0 & -1.00  & 0.50  & 0.125  &    6.6    &   5.4    &    9.8    & 1.2    \\ \hline
     0.0 & -0.53  & 0.50  & 0.250  &    2.4    &   2.3    &    4.3    & 1.2    \\ \hline
\end{tabular}
\caption{The coefficients of the expressions (\ref{xi-R-L-fitting-1})-(\ref{xi-R-L-fitting-2}), obtained by fits through the numerical values of $\xi_{R}$ and $\xi_{L}$ over the appropriate temperature ranges, 
              are reported in a few cases of interest. Here, $A_{R}^{(\pm)}$ is in units of $\sqrt{\frac{T_{c}^{R}}{2 m T_{F}}}$, $A_{L}^{(-)}$ in units of $\sqrt{\frac{T_{c}^{L}}{2 m T_{F}}}$, and $B_{R}$ in units of $T_{c}^{R}$.}
\end{center}
\end{table}
These results confirm the occurrence of a two-slope dependence for the temperature-dependent pair penetration depth (corresponding, respectively, to the full and dashed lines in 
Figs.~\ref{Figure-6} and \ref{Figure-7}), as it was anticipated in the Introduction.
In addition, these results can be regarded as assessing the quite good accuracy of our numerical calculations.

\begin{center}
{\bf E. Density profile across the interface}
\end{center}

\begin{figure}[t]
\begin{center}
\includegraphics[width=7.8cm,angle=0]{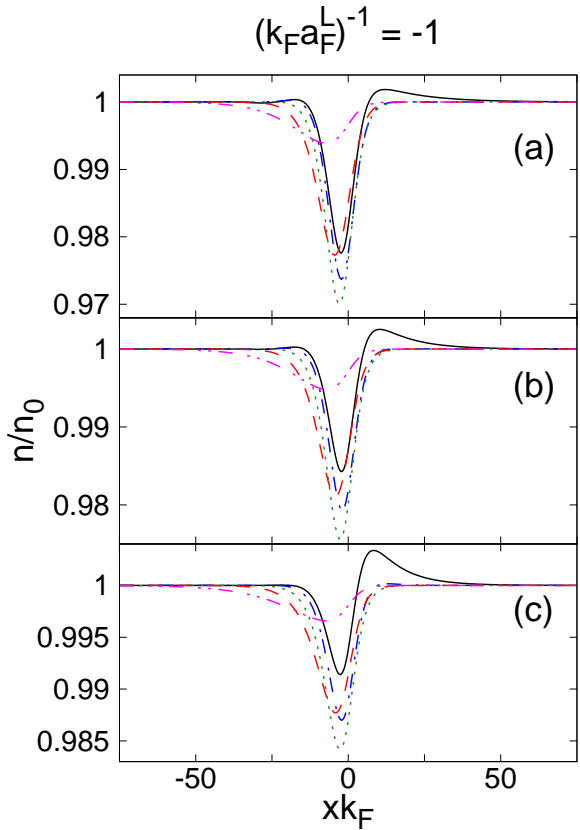}
\caption{(Color online) Density profiles (in units of the bulk density $n_{0}$) for $(k_{F} a_{F}^{L})^{-1} = -1.0$ when (a) $T_{c}^{L}/T_{c}^{R} = 8$, (b) $T_{c}^{L}/T_{c}^{R} = 4$, 
                                    and (c) $T_{c}^{L}/T_{c}^{R} = 2$.
                                    Conventions for the various curves are the same of Fig.~\ref{Figure-2}.}
\label{Figure-8}
\end{center}
\end{figure} 

As anticipated in subsection \ref{sec:proximity-NLPDA}-D, the presence of the external potential (\ref{V-model-dipole-layer}) is expected to somewhat modify the density profile near the interface, in spite of the fact that the bulk density is assumed to have the same value $n_{0}$ on both sides of the interface.
To determine the amount of this effect, we have evaluated the density profile $n(x)$ by performing the wave-vector integration in the expression (\ref{density-equation}) in spherical coordinates with the local values of 
$\Delta(x)$ and $\mu(x)$ as they vary across the interface.
The results of this calculation are shown in Fig.~\ref{Figure-8}  for $(k_{F} a_{F}^{L})^{-1} = -1.0$ and using a barrier specified by the form (\ref{definition-profile-simple}) with $k_{F} \sigma = 5.0$,
when $T_{c}^{L}/T_{c}^{R} = 8$ (upper panel), $T_{c}^{L}/T_{c}^{R} = 4$ (middle panel), and $T_{c}^{L}/T_{c}^{R} = 2$ (lower panel).
In each panel, different curves correspond to different temperatures taken between $T_{c}^{R}$ and $T_{c}^{L}$, with the same convention of Fig.~\ref{Figure-2}.
Note that, close to the interface in all cases, small (less than $3 \%$) deviations occur for $n(x)$ from its bulk value $n_{0}$.
In addition, at the lowest temperature (which in each panel is $95\%$ close to $T_{c}^{R}$ over the interval $T_{c}^{L}-T_{c}^{R}$), the depression in $n(x)$ on the left side of the interface is accompanied by a corresponding enhancement on the right side of the interface (full curves).
This dip-and-peak profile is soon washed out for increasing temperature.

\begin{figure}[t]
\begin{center}
\includegraphics[width=7.5cm,angle=0]{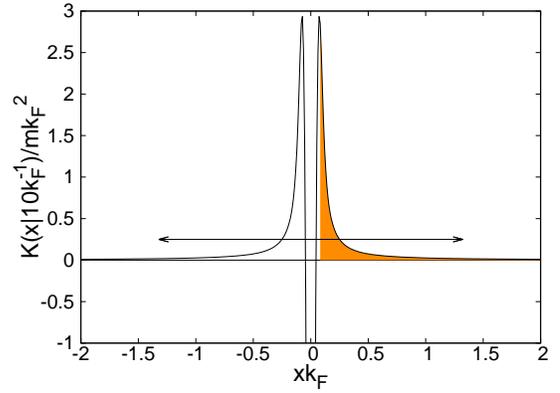}
\caption{(Color online) Typical spatial profile of the kernel $K(x|x_{0})$ of the 1D-NLPDA equation (in units of $m k_{F}^{2}$) for $(k_{F} a_{F}^{L})^{-1} = -1.0$, $T_{c}^{L}/T_{c}^{R} = 8$, 
                                     $x_{0} k_{F} = 10$, and temperature half way between $T_{c}^{R}$ and $T_{c}^{L}$.
                                     The shaded area corresponds the region over which the integral (\ref{F-function}) is calculated, while the double arrow represents the width $2 \bar{X}(x_{0})$ of the kernel 
                                     as identified by the procedure described in the text.}
\label{Figure-9}
\end{center}
\end{figure} 

\begin{center}
{\bf F. Width of the kernel of the NLPDA equation across the interface}
\end{center}

The spatial width of the kernel $K$ of the NLPDA equation (\ref{R-version-NLPDA-equation}) was shown in Ref.~\cite{Simonucci-2017} to correspond to the Cooper pair size, over the whole coupling-vs-temperature phase diagram up to the critical temperature.
This result was obtained using the values of $\Delta$ and $\mu$ that correspond to a homogeneous system for given temperature and coupling.
In the present context, however, where both $\Delta(x)$ and $\mu(x)$ vary across the interface at $x=0$ in the temperature interval $T_{c}^{R} < T < T_{c}^{L}$ of interest, the width of 
the kernel $K$ of the 1D-NLPDA equation is also expected to depend on $x$. 

To extract this dependence, we consider the kernel in real space \cite{footnote-IV} 
\begin{equation}
K(x|x_{0}) = \int_{-\infty}^{+\infty} \!\! \frac{dQ}{\pi} \, e^{2 i Q x} \, K(|Q||x_{0}) 
\label{FT-kernel}
\end{equation}
\noindent
which is an even function of $x$ and is calculated according to an expression similar to Eq.~(\ref{approximate-FT-direct}) of the Appendix.
Here, $x_{0}$ is the spatial point whereby the values of the local gap $\Delta(x_{0})$ and chemical potential $\mu(x_{0})$ enter the kernel $K(|Q||x_{0})$ in Eq.~(\ref{1D-NLPDA-equation}).
A typical profile of $K(x|x_{0})$ is shown in Fig.~\ref{Figure-9}.

The width of $K(x|x_{0})$ is then determined for given $x_{0}$ by considering the function
\begin{equation}
F(X|x_{0}) = \int_{x_{\mathrm{max}}}^{+X} \!\! dx \, K(x|x_{0}) \, ,
\label{F-function}
\end{equation}
\noindent
where $x_{\mathrm{max}} > 0$ is the position of the maximum at the right in the profile of $K(x|x_{0})$ (which also depends on $x_{0}$ and has to be determined in each case).
The function (\ref{F-function}) is found to converge asymptotically to a finite value $F(\infty|x_{0})$ when $X \rightarrow \infty$.
We thus look for the value $\bar{X}(x_{0})$ of $X$, such that $F(\bar{X}|x_{0})$ differs from $F(\infty|x_{0})$ by, say, $10 \%$.
By our definition, the width of the kernel $K$ is identified with \emph{twice} the value of $\bar{X}(x_{0})$ for any given $x_{0}$.

\begin{figure}[t]
\begin{center}
\includegraphics[width=7.8cm,angle=0]{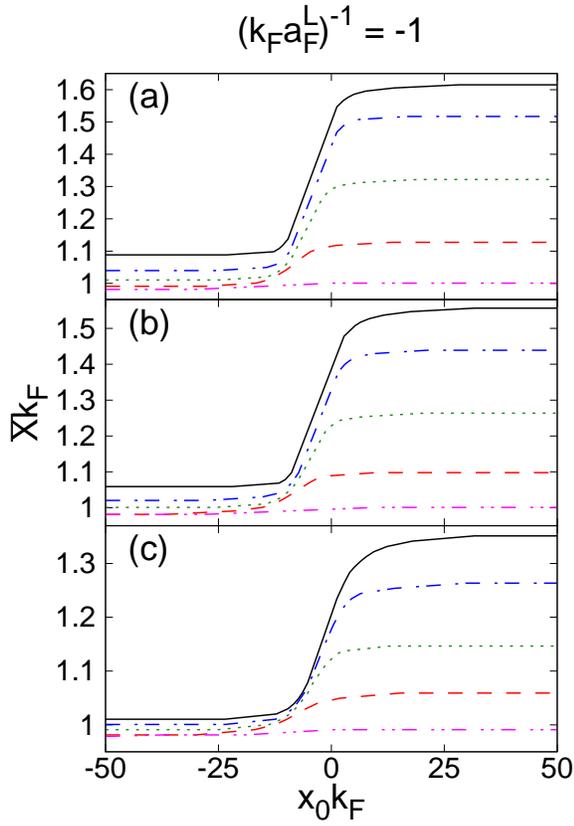}
\caption{(Color online) The quantity $\bar{X}(x_{0})$ (such that $2 \bar{X}(x_{0})$ identifies the width of the kernel of the 1D-NLPDA equation) is shown vs $x_{0}$ for $(k_{F} a_{F}^{L})^{-1} = -1.0$ 
                                     when (a) $T_{c}^{L}/T_{c}^{R} = 8$, (b) $T_{c}^{L}/T_{c}^{R} = 4$, and (c) $T_{c}^{L}/T_{c}^{R} = 2$.
                                    Conventions for the various curves are the same of Fig.~\ref{Figure-2}.}
\label{Figure-10}
\end{center}
\end{figure} 

Figure~\ref{Figure-10} shows the quantity $\bar{X}(x_{0})$ determined in this way vs $x_{0}$, for the sake of example when $(k_{F} a_{F}^{L})^{-1} = -1.0$ and $T_{c}^{L}/T_{c}^{R} = (8,4,2)$.
In each case, the various curves refer to different temperatures according to the conventions of Fig.~\ref{Figure-2}.
Note how, in each case, the shape of $\bar{X}(x_{0})$ resembles a smoothed step function, which raises from $\bar{X}(-\infty)$ to $\bar{X}(+\infty) > \bar{X}(-\infty)$ within a narrow interval
of the order of the variation of the function $G_{\sigma}(x)$ entering Eqs.~(\ref{model-function}) and (\ref{V-model-dipole-layer}).

\begin{figure}[t]
\begin{center}
\includegraphics[width=8.0cm,angle=0]{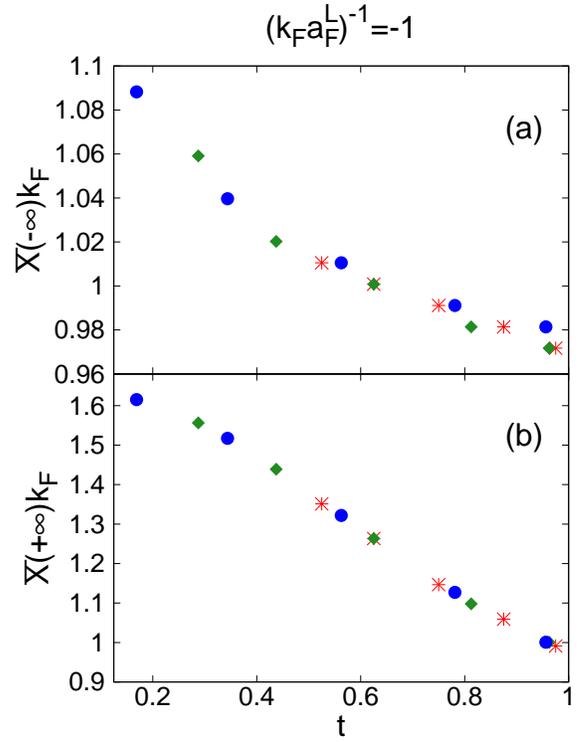}
\caption{(Color online) The values of (a) $\bar{X}(-\infty)$ and (b) $\bar{X}(+\infty)$ collected from Fig.~10 are shown vs $t=T/T_{c}^{L}$.
                                     Dots, diamonds, and stars refer to the values reported in panels (a), (b) and (c) of Fig.~10, respectively.}
\label{Figure-11}
\end{center}
\end{figure} 

In addition, in Fig.~\ref{Figure-11} we have collected the values of $\bar{X}(-\infty)$ and $\bar{X}(+\infty)$ from the three panels of Fig.~10, 
and displayed them as functions of the absolute temperature $T$.
It turns out that $\bar{X}(-\infty)$ on the extreme left and $\bar{X}(+\infty)$ on the extreme right of the interface are both decreasing functions of $T$. 
In particular, $\bar{X}(-\infty)$ decreases by about $10 \%$ when $T/T_{c}^{L}$ varies from $0.125$ to $1$, in line with what was found in Ref.~\cite{Palestini-2014} for the temperature evolution of the Cooper pair size
\emph{below} $T_{c}$ in the homogeneous case at the mean-field level.
On the other hand, $\bar{X}(+\infty)$ decreases more significantly over the same temperature interval, which would now correspond to temperatures \emph{above} $T_{c}$ in the homogeneous case,
for which it can be calculated only once pairing fluctuations beyond mean field are properly included \cite{Palestini-2014}.
Note, finally, that (twice) the values of $\bar{X}(+\infty)$ at the right of the interface are always smaller than the corresponding values of the pair penetration depth $\xi_{R}$ reported in 
Fig.~\ref{Figure-6} for the same coupling and temperature interval, thereby giving definite support to the internal consistency of the procedure we have used for identifying $\xi_{R}$.

\section{Concluding remarks} 
\label{sec:conclusions}

In this paper, we have examined theoretically the proximity effect at the interface between two superconductors with different critical temperatures under a variety of circumstances.
To the extent that the size of the Cooper pairs represents a crucial ingredient for the proximity effect, we have been able to vary this size appreciably by making the inter-particle coupling for both superconductors
to vary along the BCS-BEC crossover (although always remaining on the BCS side of unitarity, which is where the Cooper pair size is comparable with the inter-particle distance).
We have also been able to consider temperatures quite close (up to $99 \%$) to the critical temperature of either superconductor, as well as to modify the shape of the interface separating the two superconductors 
in order to assess the physical robustness of the calculations.

In this way, from the numerical profiles of the inhomogeneous gap parameter $\Delta(x)$ we have been able to extract the pair penetration depth $\xi_{R}$ on the normal (by our convention, the right) side of the interface, as a function of both coupling and temperature.
This was done in such an accurate way that the temperature dependence of $\xi_{R}$ was always found to match the behaviors expected from the analytic estimates made some time ago in Ref.~\cite{Kogan-1982} (although in that reference only for what would today be referred to as the BCS limit of the BCS-BEC crossover).
On the basis of the values attained by $\xi_{R}$ under various physical circumstances, we have also proposed a criterion for optimizing the occurrence of the proximity effect. 

All of this has been possible because the profiles $\Delta(x)$ of the gap parameter have been obtained by solving numerically the NLPDA integral equation in the form (\ref{NLPDA-eq-discretized}), instead of solving the much more demanding BdG differential equations from which the NLPDA equation was derived in Ref.~\cite{Simonucci-2014} to start with.
To this end, we have utilized the method recently provided in Ref.~\cite{Simonucci-2017} for solving the NLPDA equation, in terms of a novel efficient algorithm for calculating the Fourier transforms.
In Ref.~\cite{Simonucci-2017} it was further tested that, using the NLPDA instead of the BdG approach, not only provides a considerable gain in memory storage, but also results in a large reduction of computational time (which was there quantified in a factor of about $10^{2}$ for the case of an isolated vortex throughout the BCS-BEC crossover, for which the solution of the BdG equations is also available \cite{Simonucci-2013}).

Given the flexibility of the theoretical approach we have adopted, one may hope that the present study could stimulate a revival of the experiments that adopt similar geometry and physical arrangements, in particular by extending the work of Ref.~\cite{Polturak-1991} in a systematic way.
In addition, when a stationary current would be added to the present calculation (possibly in the presence of a sandwich of different superconductors), the local profile of the gap parameter associated with the proximity effect could be experimentally measured by tunneling spectroscopy as it was done in Ref.~\cite{Esteve-2008}.


\vspace{0.2cm}
\begin{center}
\begin{small}
{\bf ACKNOWLEDGMENTS}
\end{small}
\end{center}
\vspace{-0.1cm}

We are indebted to P. Pieri for discussions about the choice (\ref{V-model-dipole-layer}) of the external potential.

\appendix   
\section{METHOD FOR THE NUMERICAL SOLUTION OF THE 1D-NLPDA EQUATION}
\label{sec:appendix-A}
\vspace{-0.2cm}

An efficient method to solve numerically the ($\mathbf{Q}$-version (\ref{Q-version-NLPDA-equation}) of the) NLPDA equation was set up in Appendix B of Ref.~\cite{Simonucci-2017} in any dimension D.
The method rests on the peculiar properties of the Fourier transform of the wave functions of the D-dimensional harmonic oscillator, when expressed in terms of the generalised Laguerre polynomials for
a gap parameter $\Delta(\mathbf{r})$ with a given spatial symmetry.

For the 1D geometry of the proximity effect of interest to the present paper, where the gap parameter $\Delta(x)$ has both even and odd components in $x \leftrightarrow -x$, one could thus split
into even and odd components both the coupling parameter $g(x)$ of Eq.~(\ref{model-function}) and the product $g(x) \Delta(x)$ which appears on the left-hand side of Eq.~(\ref{1D-NLPDA-equation}), to end up with two coupled equations for the even and odd components of $\Delta(x)$.
Given the simple 1D geometry of interest, however, one can most simply rephrase the method developed in Appendix B of Ref.~\cite{Simonucci-2017} in terms of Hermite polynomials instead of generalised Laguerre polynomials.
For completeness, in the following we shall concisely report the relevant expressions needed to solve numerically Eq.~(\ref{1D-NLPDA-equation}), which are obtained by rephrasing in terms of Hermite polynomials the essential steps described in Appendix B of Ref.~\cite{Simonucci-2017}, to which we refer the reader for additional details.

Consider a 1D harmonic oscillator with mass $m=2 \lambda^{2}$ and frequency $\omega = 1$ (with the parameter $\lambda$ introduced to give additional flexibility to the numerical calculations). 
Its (normalized) eigenfunctions have the form:
\begin{equation}
\psi_{n}(x) = (2 \lambda^{2})^{1/4} \, e^{- \lambda^{2} x^{2}} \mathcal{H}_{n}(\sqrt{2} \lambda x) 
\label{eigenfunction-x-space}
\end{equation}
\noindent
with 
\begin{equation}
\mathcal{H}_{n}(x) = \frac{1}{\pi^{1/4} \sqrt{2^{n} n !}} \,\, H_{n}(x)
\label{modified-Hermite-polynomials}
\end{equation} 
\noindent
where $H_{n}(x) \, (n=0,1,\cdots)$ are Hermite polynomials, such that 
\begin{equation}
\int_{-\infty}^{+\infty} \!\! dx \, e^{-x^{2}} \, \mathcal{H}_{n}(x) \, \mathcal{H}_{n'}(x) = \delta_{nn'} \, .
\label{normalization}
\end{equation}
The corresponding Fourier transform is given by
\begin{eqnarray}
\tilde{\psi}_{n}(Q) & = & \int_{-\infty}^{+\infty} \!\! dx \, e^{-2iQx} \, \psi_{n}(x) 
\nonumber \\
& = & (-i)^{n} \, \left(\! \frac{2 \pi^{2}}{\lambda^{2}} \!\right)^{1/4} e^{- Q^{2}/\lambda^{2}} \mathcal{H}_{n}\left(\! \frac{\sqrt{2} Q}{\lambda} \!\right) 
\label{eigenfunction-Q-space}
\end{eqnarray}
\noindent
(for clarity, in this Appendix we add a tilde to the symbol of the Fourier transform).

Owing to the property of the Fourier transforms
\begin{equation}
\int_{-\infty}^{+\infty} \!\! dx \ \psi_{n}(x) \Delta(x) = \int_{-\infty}^{+\infty} \!\! \frac{dQ}{\pi} \ \tilde{\psi}_{n}^{\ast}(Q) \tilde{\Delta}(Q) \, ,
\label{property-FT}
\end{equation}
\noindent
we can write in terms of the expressions (\ref{eigenfunction-x-space}) and (\ref{eigenfunction-Q-space}):
\begin{eqnarray}
& & \int_{-\infty}^{+\infty} \!\! dx \, e^{-x^{2}} \, \mathcal{H}_{n}(x) \, e^{x^{2}/2} \, \Delta \left(\! \frac{x}{\sqrt{2} \lambda} \,\right)
\nonumber \\
& = & \frac{i^{n} \lambda}{\sqrt{\pi}} \! \int_{-\infty}^{+\infty} \!\! dx \, e^{-x^{2}} \, \mathcal{H}_{n}(x) \, e^{x^{2}/2} \,  \tilde{\Delta} \left(\! \frac{\lambda x}{\sqrt{2}} \,\right) \, .
\label{equality-x_vs_Q}
\end{eqnarray}

The above expressions can be cast in an approximate form useful for numerical calculations, by introducing a Gaussian quadrature of the form (cf. Eq.~\ref{normalization})):
\begin{equation}
\int_{-\infty}^{+\infty} \!\! dx \, e^{-x^{2}} \, \mathcal{H}_{n}(x) \, \mathcal{H}_{n'}(x)
= \sum_{j=1}^{N} w_{j} \, \mathcal{H}_{n}(x_{j}) \, \mathcal{H}_{n'}(x_{j}) = \delta_{nn'} 
\label{Gaussian-quadrature}
\end{equation}
\noindent
where the points $\{ x_{j};j=1,\cdots,N \}$ and the (positive definite) weights $\{ w_{j};j=1,\cdots,N \}$ have to be determined.
We then write for the left-hand side of Eq.~(\ref{equality-x_vs_Q})
\begin{equation}
\int_{-\infty}^{+\infty} \!\! dx \, e^{-x^{2}} \, \mathcal{H}_{n}(x) \, e^{x^{2}/2} \, \Delta \left(\! \frac{x}{\sqrt{2} \lambda} \! \right) \simeq \sum_{j=1}^{N} S_{nj} \, y_{j} \Delta \left(\! \frac{x_{j}}{\sqrt{2} \lambda} \! \right)
\label{approximate-equality-x}
\end{equation}
\noindent
as well as for the right-hand side of Eq.~(\ref{equality-x_vs_Q})
\begin{equation}
\int_{-\infty}^{+\infty} \!\! dx \, e^{-x^{2}} \, \mathcal{H}_{n}(x) \, e^{x^{2}/2} \, \tilde{\Delta} \left(\! \frac{\lambda x}{\sqrt{2}} \,\right) \simeq \sum_{j=1}^{N} S_{nj} \, y_{j} \tilde{\Delta} \left(\! \frac{\lambda x_{j}}{\sqrt{2}} \! \right)
\label{approximate-equality-Q}
\end{equation}
\noindent
where we have introduced the quantities
\begin{equation}
S_{nj} = \mathcal{H}_{n}(x_{j}) \, \sqrt{w_{j}}  \,\,\,\,\,\, , \,\,\,\,\,\, y_{j} = e^{x_{j}^{2}/2} \, \sqrt{w_{j}} \,\,\,\,\,\, , 
\label{definition-matrix_S-y}
\end{equation}
\noindent
such that
\begin{equation}
\sum_{j=1}^{N} S_{nj} \, S^{T}_{jn'} = \delta_{n n'} \,\,\,\,\,\, , \,\,\,\,\,\, \sum_{n=0}^{N-1} S^{T}_{jn} \, S_{nj'} = \delta_{j j'} \, .
\label{S-vs-S^T}
\end{equation}
\noindent
Entering the results (\ref{approximate-equality-x}) and (\ref{approximate-equality-Q}) into Eq.~(\ref{equality-x_vs_Q}), we obtain approximately
\begin{equation}
\sum_{j=1}^{N} S_{nj} \, y_{j} \Delta \left(\! \frac{x_{j}}{\sqrt{2} \lambda} \! \right) = \frac{i^{n} \lambda}{\sqrt{\pi}} \, \sum_{j=1}^{N} S_{nj} \, y_{j} \tilde{\Delta} \left(\! \frac{\lambda x_{j}}{\sqrt{2}} \! \right) \, ,
\label{approximate-equality-x_vs_Q}
\end{equation}
\noindent
from which we can extract alternatively
\begin{equation}
\Delta \left(\! \frac{x_{j}}{\sqrt{2} \lambda} \! \right) = \frac{\lambda}{\sqrt{\pi} \, y_{j}} \, \sum_{n=0}^{N-1} \sum_{j=1}^{N} \, i^{n} \, S^{T}_{jn} \, S_{nj'} \, y_{j'} \, \tilde{\Delta} \left(\! \frac{\lambda x_{j'}}{\sqrt{2}} \! \right) 
\label{approximate-FT-direct}
\end{equation}
\noindent
and
\begin{equation}
\tilde{\Delta} \left(\! \frac{\lambda x_{j}}{\sqrt{2}} \! \right)  = \frac{\sqrt{\pi}}{\lambda \, y_{j}} \, \sum_{n=0}^{N-1} \sum_{j=1}^{N} \, (-i)^{n} \, S^{T}_{jn} \, S_{nj'} \, y_{j'} \, \Delta \left(\! \frac{x_{j'}}{\sqrt{2} \lambda} \! \right)
\label{approximate-FT-inverse}
\end{equation}
\noindent
where $\frac{x_{j}}{\sqrt{2} \lambda}$ refers to values of $x$ and $\frac{\lambda x_{j}}{\sqrt{2}}$ to values of $Q$, with the two meshes of $x$ and $Q$ points closely interlinked to each other.
In these expressions, both the number of points $N$ in the two meshes and the parameter $\lambda$ can be varied to achieve optimal convergence of the Fourier transform from $\Delta(x)$ to $\Delta(Q)$ (and vice versa).
The two results (\ref{approximate-FT-direct}) and (\ref{approximate-FT-inverse}) taken together provide an efficient algorithm to calculate the Fourier transform of any function in 1D.

There remains to determine the sets of points $\{x_{j}\}$ and the corresponding weights $\{w_{j}\}$ that appear in the definitions (\ref{definition-matrix_S-y}).
To this end, we take advantage of the recursion relation \cite{MOS-1966} 
\begin{equation}
\sqrt{n+1} \, \mathcal{H}_{n+1}(x) - \sqrt{2} \, x \, \mathcal{H}_{n}(x) + \sqrt{n} \, \mathcal{H}_{n}(x) = 0
\label{recursion-relation}
\end{equation}
\noindent
which we apply recursively from $n=0$ to $n=N-1$, and choose for $x$ the $N$ values $\bar{x}$ such that $\mathcal{H}_{N}(\bar{x}) = 0$.
In this way, we end up with the following $N \times N$ eigenvalue problem:
\begin{widetext}
\begin{equation}
\left(
\begin{array}{cccccc}
- \sqrt{2} \, \bar{x} & 1  & 0 & \cdots & \\
1 & - \sqrt{2} \, \bar{x} & \sqrt{2} & 0 & \cdots \\
0 & \sqrt{2} & - \sqrt{2} \, \bar{x} & \sqrt{3} & 0 & \cdots \\
\cdots & \cdots & \cdots & \cdots & \cdots & \cdots \\
 & \cdots & 0 & \sqrt{N-2} & - \sqrt{2} \, \bar{x} & \sqrt{N-1} \\
 &  & \cdots & 0 \, & \sqrt{N-1} & - \sqrt{2} \, \bar{x} 
\end{array}
\right)
\left(
\begin{array}{c}
\mathcal{H}_{0}(\bar{x}) \\
\mathcal{H}_{1}(\bar{x}) \\
\mathcal{H}_{2}(\bar{x}) \\
\cdots \\
\mathcal{H}_{N-2}(\bar{x}) \\
\mathcal{H}_{N-1}(\bar{x})
\end{array}
\right)
=
\left(
\begin{array}{c}
0\\
0 \\
0 \\
\cdots \\
0 \\
0
\end{array}
\right) \, .
\label{big-eigenvalue-problem}
\end{equation}
\noindent
By diagonalizing the real and symmetric matrix on the left-hand side of Eq.~(\ref{big-eigenvalue-problem}), we obtain eventually the $N$ (distinct) eigenvalues $\bar{x}_{j}$ (with $j=1,2,\cdots,N$) and the corresponding $N$ eigenvectors \begin{small} $\left(\mathcal{H}_{0}(\bar{x}_{j}),\mathcal{H}_{1}(\bar{x}_{j}),\mathcal{H}_{2}(\bar{x}_{j}),\cdots,\mathcal{H}_{N-2}(\bar{x}_{j}),\mathcal{H}_{N-1}(\bar{x}_{j}) \right)$\end{small},
whose normalization condition 
\end{widetext}
\begin{equation}
\sum_{n=0}^{N-1} \mathcal{H}_{n}(\bar{x}_{j}) \, \mathcal{H}_{n}(\bar{x}_{j}) = \frac{\delta_{j j'}}{w_{j}}
\label{normalization-condition}
\end{equation}
\noindent
provides the weights $w_{j}$ according to the second identity in Eq.(\ref{S-vs-S^T}).

The above results can be used to solve the 1D-NLPDA integral equation (\ref{1D-NLPDA-equation}) with variable coupling constant $g(x)$ in an efficient way.
The ensuing discretized form of this integral equation is reported in Eq.~(\ref{NLPDA-eq-discretized}) of the main text.


\newpage

\end{document}